\documentclass[5p]{elsarticle}

\usepackage{lineno,hyperref}
\modulolinenumbers[5]

\usepackage{amssymb,amsmath}
\usepackage{graphicx}
\usepackage{dcolumn}
\usepackage{bm}

\def\e{\begin{equation}}
\def\f{\end{equation}}

\usepackage{enumerate}

\begin{document}

\begin{frontmatter}

\title{Double-resonant decoupling method in very dense dipole arrays}

\author[aalto]{M. S. M. Mollaei \corref{cor}}
\author[itmo]{A. Hurshkainen}
\author[itmo]{S. Kurdjumov}
\author[aalto]{C. Simovski}
\address[aalto]{Aalto University, School of Electrical Engineering, Department of Radio Science and Engineering, P.O. Box 13000, 00076 Aalto, Finland}
\address[itmo]{ITMO University, Kronverkskiy pr. 49, 197101, St. Petersburg, Russia}
\cortext[cor]{Corresponding author at: Department of Electronics and Nanoengineering,
Aalto University, P.O. 13000, FI-00076, Finland. Email address: masoud.2.sharifianmazraehmollaei@aalto.fi}


\begin{abstract}
In this paper an approach for broadening of operational band in a dense array of dipole antennas by implementing passive split-loop resonators (SLRs) as decouplers is presented. Compared to the previous method, where three closely located active dipoles were decoupled by two passive dipole, the operational band is significantly improved from $0.5 \%$ to $1.6 \%$ at the same level of decoupling -8dB for the cross-talk and inter-channel transmittance. To delineate, the presence of two SLRs results in birefringence of the resonant interaction of SLRs which creates two different eigenmodes for decoupling. As a result, a dual-resonant decoupled band is obtained. Alongside with analytical investigation, numerical and experimental investigations verify the veracity of our approach. Moreover, the possibility of decoupling by SLRs for arrays with more active dipoles is investigated numerically.

\end{abstract}

\begin{keyword}
Active metasurface, Antenna array, Decoupling, Split loop resonator.
\end{keyword}

\end{frontmatter}

\section{Introduction}
\label{sec:intro}
Array antennas have made focus of research for decades due to their practical features such as gain enhancing, beam steering and path diversity giving. Because of these features, array antennas have been intensively used in many applications, namely, radio astronomy, radio frequency identification (RFID), multi-input multi-output (MIMO) and magnetic resonance imaging (MRI) \cite{1,2,3}. Yet, coupling between array elements is a critical impasse especially when the distance between them is less than $\lambda / 10$, $\lambda $ is operation wavelength.

Former, many techniques for solving this problem have been presented. Most practical ones are: using adaptive active circuitry for compensating induced electromotive forced over each element created by the other elements; using absorbing sheets between elements to isolate them electromagnetically; using electromagnetic band gap structures (three is the minimal number of cells that must be located between elements) \cite{4,5,6,7}. However, these approaches are, respectively, too expensive for commercial applications; destructive for antenna pattern and efficiency; and applicable when the distance between elements is larger than $\lambda / 12$.  
For the case that array is constituted of loop antennas, overlapping elements and using passive loops are most common method \cite{8}. 
Some passive loops have also been used for dipole antennas decoupling \cite{9}.
In the case of dipole arrays, the most critical situation raises when the distance between element is smaller than $\lambda / 20$. This is the case in ultra-high field MRI when the distance between dipoles is $d=\lambda / 30$ (operation frequency is the Larmor frequency 300 MHz for 7 Tesla MRI scanners).  
Lau and et al. have presented a method for decoupling two closely located dipoles; which proves the need of active load for their complete decoupling \cite{10}. Moreover, their approach has not been discussed for resonant dipoles.

In our previous works, first we introduced a method of passive decoupling between two and three closely resonant dipoles ($d=\lambda / 33 = 30$ mm) by adding one and two the same but passive dipoles between them, respectively \cite{11}. Using this approach, transmittance coefficient between active dipoles has been decreased from -2dB (for the reference case -- without passive dipoles) to -13 dB while the relative bandwidth was 0.5\% at the level of -8 dB. Then in work \cite{12}, we have enhanced decoupling bandwidth between two active dipoles through replacing the passive dipole of \cite{11} with a split-loop resonator (SLR). However, no one has used two and more SLRs to decouple dense arrays of dipole antennas. 

In this paper, we aim to enhance operation band of dense dipole array through replacing two passive dipoles in \cite{11} with the two passive SLRs. The possibility of decoupling in the dense array is discussed analytically, numerically and experimentally in next sections. In fact, interaction between these two passive SLRs results in two eigenmodes for decoupling which widen operational band of the system. Finally, the possibility of extending this method for arrays with more active dipoles is numerically investigated.
\section{Analytical Model}\label{sec2}

\begin{figure}
	\centering{\includegraphics[width=70mm]{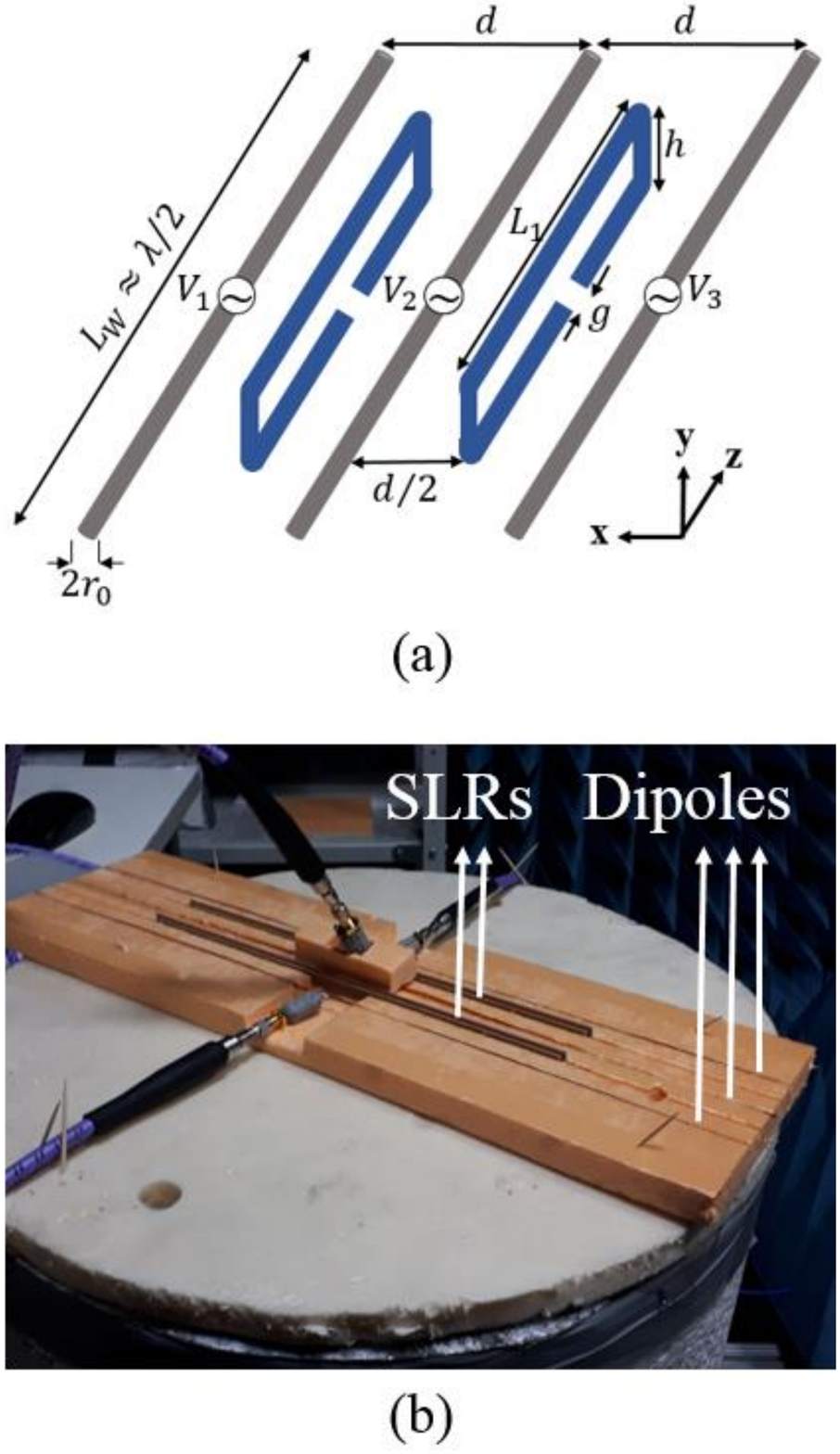}}
	\caption{Array of three dipole antennas decoupled by two SLRs. (a) Schematic view, (b) fabricated prototype.}\label{fig1}
\end{figure}

Consider an array formed by three active dipoles and two passive SLRs located as shown in figure 1(a). Let us follow to the logics of our work \cite{11} where the general condition of the decoupling was obtained in terms of the self- and mutual impedances and the geometry of the decoupling scatterers was not specified. For an array of three active dipoles and two passive scatterers we have (see also in figure \ref{fig1}(a)):
	\begin{equation}
	{Z''}(Z_0 + Z_{00}) = \left(Z'\right)^2 + Z'Z''',  \label{cn1}
	\end{equation}
	\begin{equation}
	{{{Z_{00}}\left[ {{{\left({Z'} \right)}^2} + {{\left( {Z''''} \right)}^2}} \right] - 2{Z_0}Z'Z''}} =\left(Z_{00}^2 - Z_0^2\right){Z''''}.  \label{cn2}
	\end{equation}
Here $Z_0$ and $Z_{00}$ are, respectively:
\begin{itemize}
	\item
	self-impedance of our passive scatterers,
	\item
	mutual impedance of two passive scatterers.
\end{itemize}

Next, $Z' $, $Z'' $, $Z''' $ and $Z'''' $ are mutual impedances, respectively:
\begin{itemize}
	\item
	of two adjacent elements: an active dipole and an SLR,
	\item
	of two neighboring dipoles (1 and 2, 2 and 3),
	\item
	of two distant elements, one active and one passive (left dipole and right SLR and vice versa),
	\item
	of two distant dipoles (1 and 3).
\end{itemize}

It can be shown analytically that the absolute values of $Z_0 $ and $Z' $ are much larger than those of $Z''' $ and $Z_{00}$ \cite{11,13}.
Therefore, we can use the following approximation for decoupling the neighboring dipoles instead of (\ref{cn1}):
\begin{equation}
Z_0 Z^{''} = {(Z^{'})}^2 .  \label{cn3}
\end{equation}
Considering equations (\ref{cn2}) and (\ref{cn3}), any type of scatterers which satisfies these equations results in the decoupling between all three array elements.
In \cite{11}, these decoupling conditions were satisfied by two passive dipoles identical to the active dipoles but shortcut at the center.
It is possible to prove that replacing these passive dipoles by passive SLRs  allows to satisfy these equations as well \cite{12}. However, equation
(\ref{cn2}) cannot be for resonant SLRs reduced to the same form (\ref{cn3}), like it was done in \cite{11} for resonant dipoles.
In the case of passive dipoles, the enlargement of the array from 2 to 3 active elements and from 1 to 2 passive elements does not
imply anything new for decoupling.

In the case of SLRs it is not so. Decoupling condition for the adjacent dipoles (3) is analytically solved in \cite{12} and the condition is satisfied at the frequency $\omega \approx 1.0432{\omega _0}$ (proper dimensions for SLR to decouple two dipoles are calculated based on this analytical solution). Decoupling condition for the distant dipoles (2) is numerically solved. There is no analytical solution for mutual coupling between two SLRs in the literature. In fact, there are two components of their mutual impedance resonating at nearly the same frequency. First one is the electric-dipole component related with the symmetric part of the loop current (the same current in the top and bottom sides of the SLR). Second one is the magnetic-dipole component, related to the asymmetric part of the SLR current. For electric-dipole component the problem was solved in [12]. For the magnetic-dipole component the problem can be solved in principle also analytically. However, calculating  it is out of the scope of this paper. At the highest frequency ($\omega \approx 1.0432{\omega _0}$), that corresponds to the best decoupling of the neighboring dipoles, the eignemode of two SLRs is polarized in phase. At the middle frequency ($\omega \approx 1.017{\omega _0}$), the eigenmode of two SLRs still remains polarized in phase (thet will be numerically demonstrated in the next section). At the lowest frequency ($\omega \approx 0.982{\omega _0}$), the eigenmode of two SLRs is polarized with opposite phase.

In the case simulated below $\omega_0=2\pi f_0$, $f_0=284$ MHz. It is the common resonant frequency of the active dipoles and SLRs in our array.
All three decoupling frequencies are within the resonance band of the dipole antennas.
This is concise with the initial assumption of the analytical model and promises us the suitable antenna efficiency (not damaged by the decoupling elements).
Additionally, it means that the interaction between two SLRs grants us a broader operational band. Really, the maximal decoupling of the neighboring and distant dipoles
hold at three very close frequencies, i.e. the bands of the decoupling overlap and a total decoupling band is broader than that in the case of one decoupling SLR.
In the broadened operational band the decoupling will be though not ideal but suitable. And the decoupling band for the distant dipoles 1 and 3 should manifest two local minima
of $S_{13}$.

In the next section, the accuracy of our analytical approach is numerically and experimentally verified. Moreover, the possibility of extending this method for an array constituted of four active dipoles, decoupled by three SLRs is numerically proved. 
	
\section{Numerical and Experimental Verifications}\label{sec4}

\begin{figure}
	\centering{\includegraphics[width=85mm]{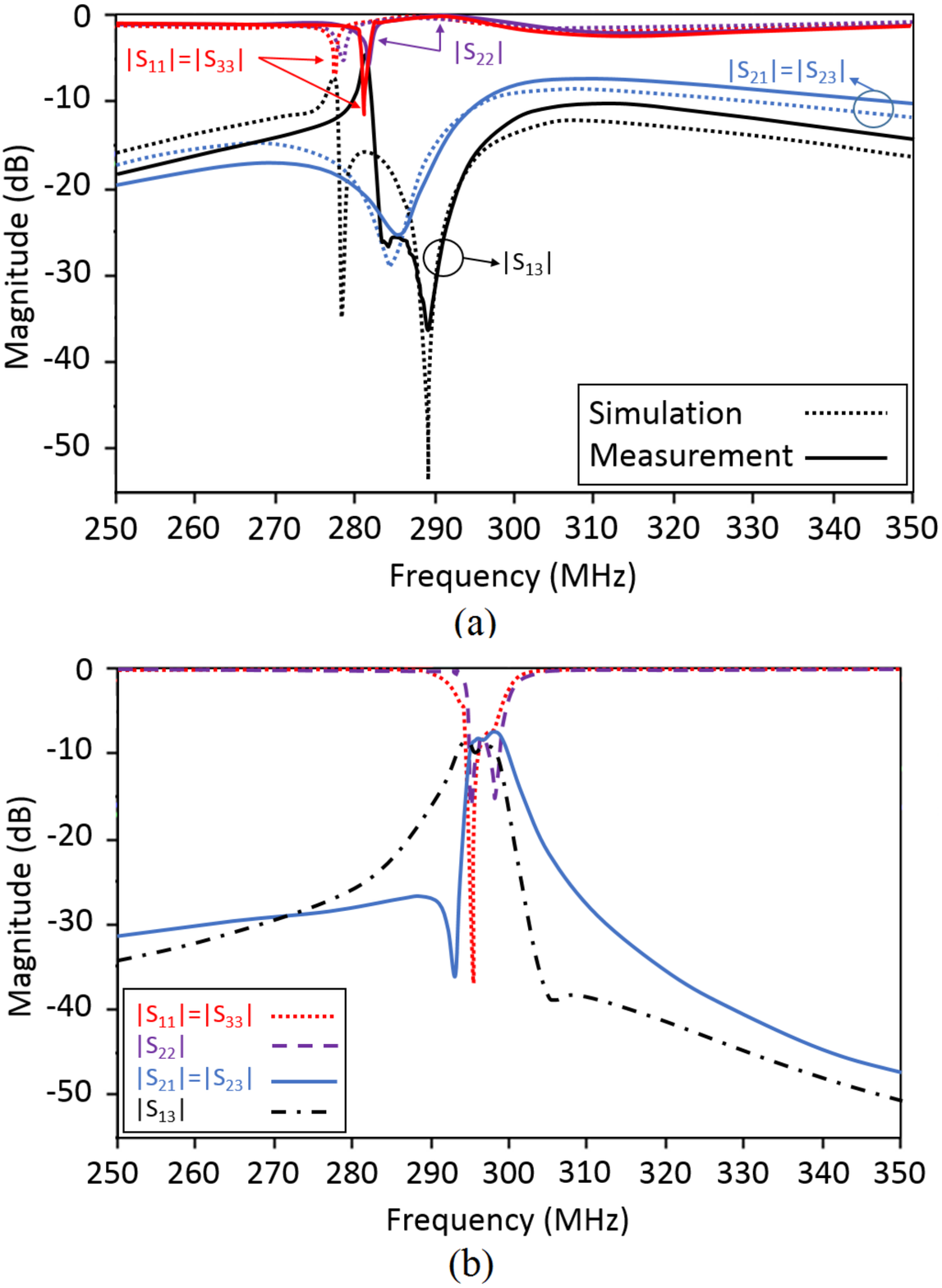}}
	\caption{Simulation and measurement performance results of the structure consisting of three active dipoles and two SLRs. (a) Mismatched regime, (b) matched regime.} \label{fig3}
\end{figure}
We simulated the structure depicted in figure 1(a) using CST Microwave Studio and experimentally verified our simulations
measuring S-parameters of a fabricated prototype shown in figure~\ref{fig1}(b). Similar to \cite{11}, simulation has been carried out for mismatched and matched regimes.
Having in mind the application of our theory for ultra-high field MRI (7T), we designed the structure so that it would operate at frequencies between 270 MHz and 300 MHz.

For the reference case (no SLRs) coupling coefficients for the adjacent dipoles
at the resonant frequency $f_0$ in the mismatched and matched regimes are nearly equal $-5$ dB and $-2$ dB, respectively.
Introducing 2 SLRs with following geometric parameters \textbf{(based on analytical solution in \cite{12})} -- $L_1 =$ 322 mm, $h =$ 10 mm, $g =$ 30 mm, $r_0 =$ 1 mm, and $d =$ 30 mm -- between the dipoles
as shown in figure 1(a), we come to the results presented in figure~\ref{fig3}.

The measurements were done similarly to the corresponding measurements for passive dipoles \cite{11}.
 Dipoles and SLRs are made of copper wire with the same dimensions used in the simulation. The feeding points of the dipoles are soldered to connectors through which dipoles are connected to a Vector Network Analyzer.
As well as in \cite{11} for simplicity of the experiment, we avoided the fabrication of the matching circuit, because the $S$-parameters obtained for the mismatched regime can be recalculated to the
regime of the perfect matching using the existing software.

Simulation and measurement results for mismatched case are seen in figure~\ref{fig3}(a). This regime is important because the coupling coefficients manifest the aforementioned minima
in the band of decoupling. Moreover, at the individual resonance the decoupling in the mismatched regime grants the self-tuning -- minima of the reflection coefficients $S_{22}$ (central dipole) and
$S_{11}=S_{33}$ (edge dipoles) are clearly seen in figure~\ref{fig3}(a). 
This self-tuning is a feature of antennas decoupling. When they are strongly coupled the most part of the energy is transferred between them in a non-radiative way. 
Since the antennas are connected to the voltage sources with the internal impedance 50 Ohm, and when they are coupled the radiation is absorbed in these resistors and   
the radiation resistance is damped by the coupling. The coupling of our antennas is overcritical -- they radiate very weakly even within the band of their individual resonance. 
At a frequency of decoupling the radiation resistance of every antenna sharply grows. 
Instead of being absorbed in the resistors connected to two other antennas, the power is radiated to free space that grants the sharply resonant increase 
of the radiation resistance and, therefore, the self-tuning. Of course the decoupled antennas are strongly coupled to the SLRs, and this coupling is also destructive 
for their radiation. This is why the self-tuning minima of the reflectance are small. However, they are noticeable because the passive scatterers are practically lossless
and cannot absorb the power transferred by near fields. 

The simulated and measured curves for the reflection coefficients almost coincide. The experimental and theoretical curves for the coupling coefficients 
slightly differ from one another, but are in a qualitative agreement. What is important -- in the experimental curve for $S_{13}$ we clearly distinguish two minima predicted by the theory and very pronounced in the simulated curve. The difference between the simulations are measurements is here standard -- more smooth and spread experimental curve implies that some 
electromagnetic losses present in the experimental setup were not taken into account in simulations. The minima for two coefficients $S_{13}=S_{31}$ and $S_{12}=S_{21}=S_{23}=S_{32}$ strongly overlap and grant the broad decoupling band of width nearly 5.5\% on the level -10 dB, from 281.6 to 297.3 MHz. 

\begin{figure}
	\centering{\includegraphics [width=85 mm]{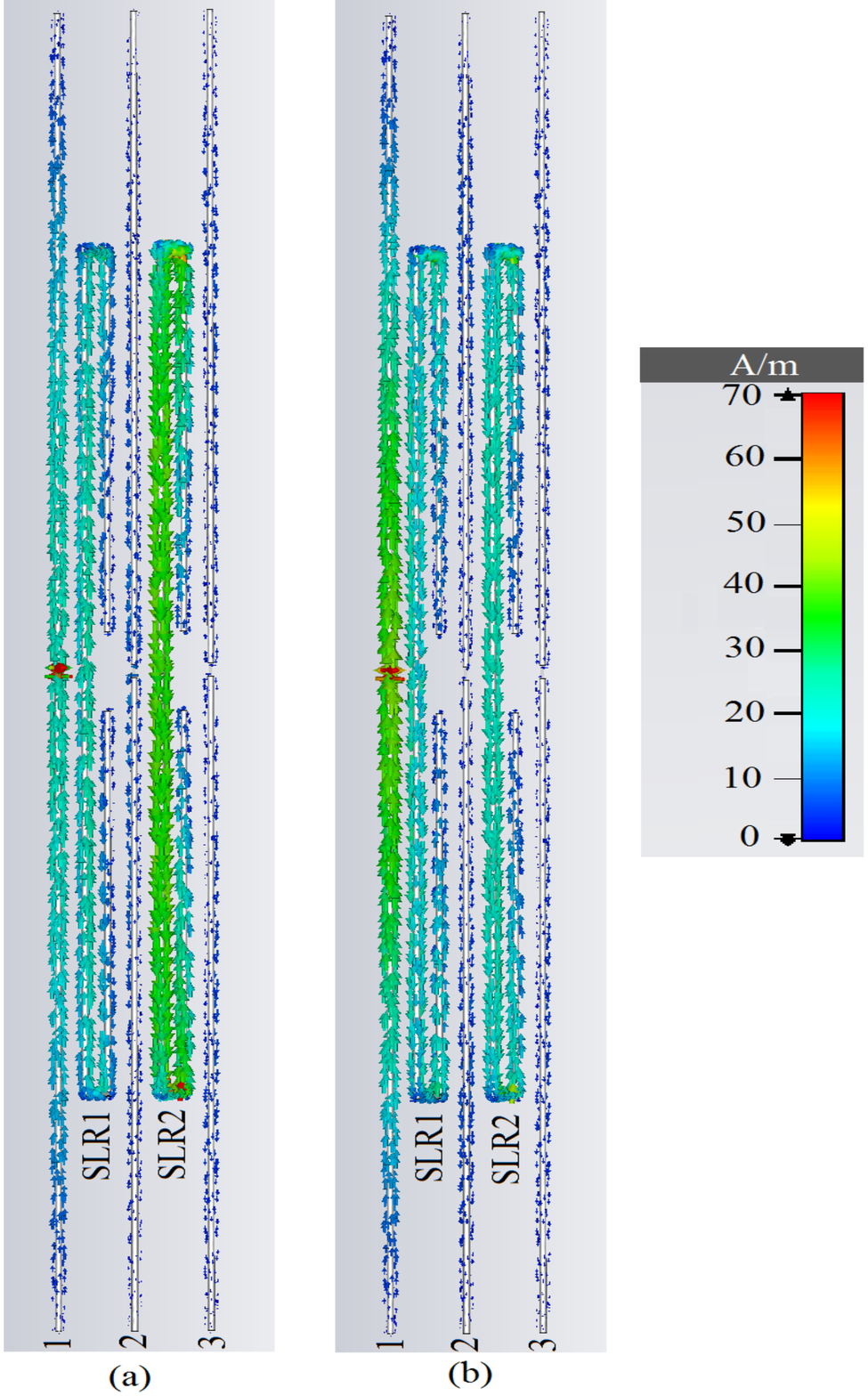}}
	\caption{Induced current over SLRs. (a) Induced out of phase currents over SLRs in 278.42 MHz, (b) induced in phase currents over SLRs in 289.39 MHz.}\label{fig4}
\end{figure}

\begin{figure}
	\centering{\includegraphics [width=97 mm]{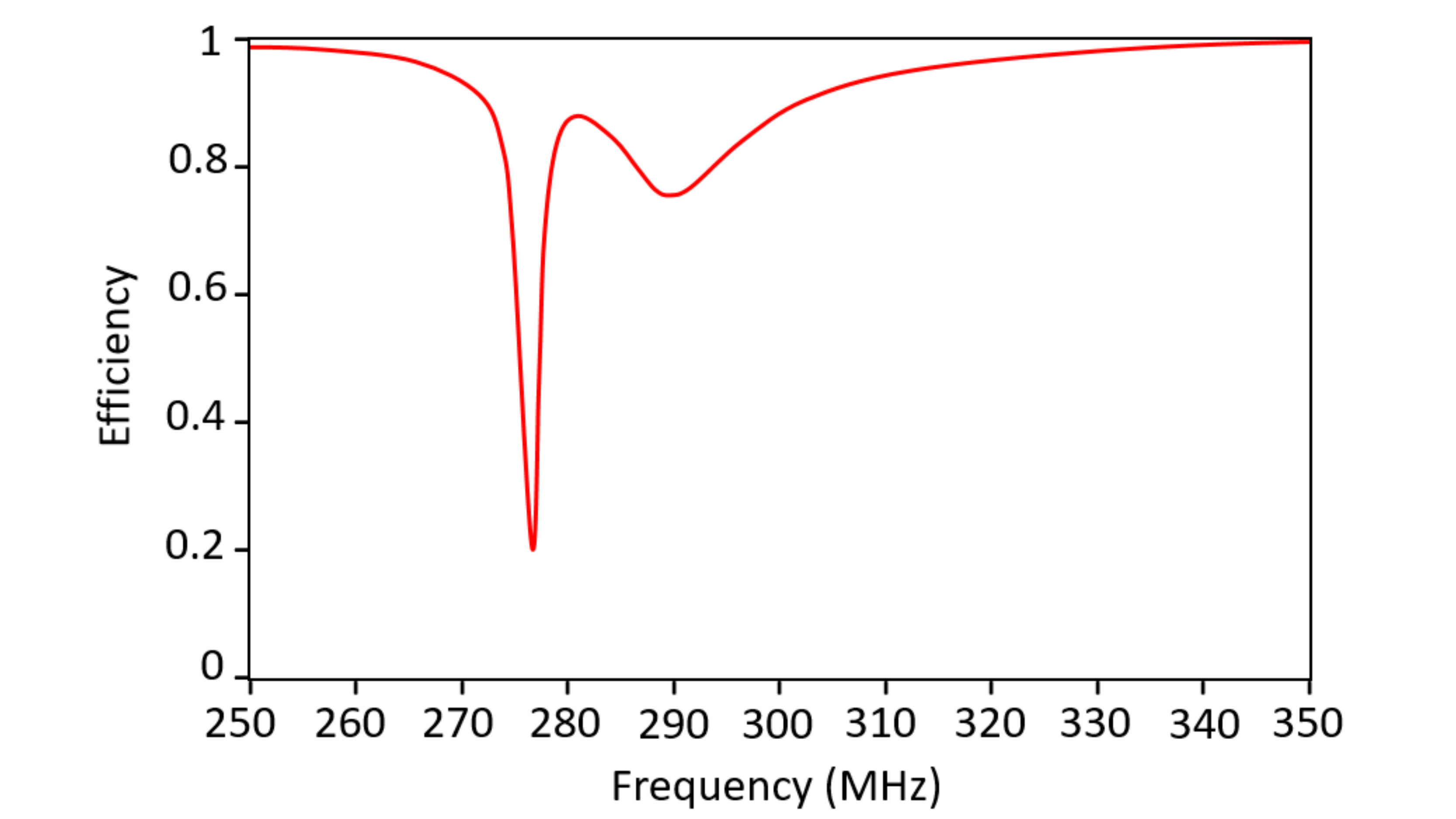}}
	\caption{Radiation efficiency of the structure.}\label{ad2}
\end{figure}

In the matched regime a similarly broad band is manifested as a band of matching in which the maximum of $S_{12}$ and $S_{13}$ is suppressed from -2 dB to -8 dB. 
Here the criterion is taken on the level -8 dB, since higher decoupling in the matched regime is impossible and the coupling below the critical value of -6 dB is acceptable.   
The same refers to the reflection coefficient of the antenna. 
The results obtained by adding a virtual lossless matching circuitry between the voltage sources and the dipole antennas are shown in figure~\ref{fig3}(b). 
Since the same procedure can be applied to both simulations and experimental results, we show here only the simulated curves -- the same 
agreement between the experimental and numerical data as in figure~\ref{fig3}(a) keeps for the matched case. According to figure~\ref{fig3}(b), 
the operation bandwidth of the antenna array is enhanced due to the replacement of the passive dipoles by passive SLRs
from $0.5\%$ to $1.6\%$ (if we apply to the results for coupling coefficients from \cite{11} the same criterion of -8 dB) from 294 to 298.6 MHz (matched in a way to be practical for 7T MRI scanner). 

Further, to prove the veracity of our expectation, we monitored the induced currents over SLRs in frequencies coincide with the minima of $S_{13} = S_{31}$ ( 278.42 and 289.39 MHz). Figure \ref{fig4} shows induced currents over SLRs in these frequencies. As we expected, in 278.42 MHz the currents induced over SLRs are out of phase and the currents induced over them in 289.39 MHz are in phase, while in both frequencies these modes decouple antennas 2 and 3 from antenna 1. Besides, efficiency of the structure is shown in figure~\ref{ad2}.

\begin{figure}
	\centering{\includegraphics[width=98mm]{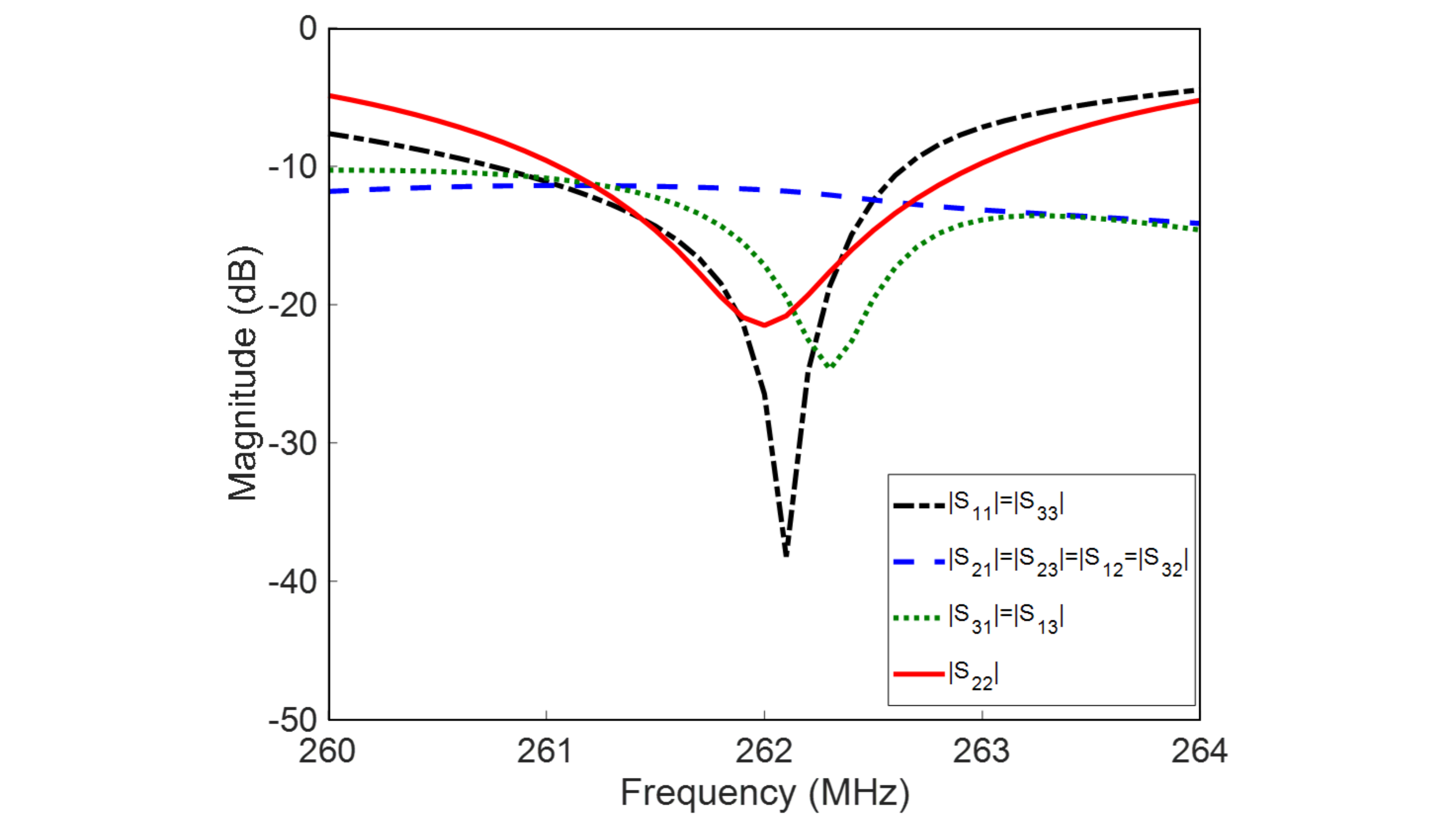}}
	\caption{S-parameters of the structure in the presence of the phantom in the matched regime.}\label{fig7}
\end{figure}

Needless to say that considering main application of the work -- MRI -- when a high permittivity object such as human body is located beneath the structure, the decoupling will be enhanced to -10 dB.
In the practical case when human body is located beneath the structure, decoupling level will enhanced enough due to high loading by the body. In fact, in the presence of the body, most of radiated energy will be propagated in the body. To clarify more, we simulated a case when a cubic phantom with permittivity 78 and conductivity 1.56 (parameters are chosen according to human body) is located beneath the structure. In this case, SLRs should shift up a bit to compensate asymmetry of the structure. Below, simulation result in the matched case in the presence of the phantom is shown in figure~\ref{fig7}. In this case, the decoupling level enhanced to -10 dB in the presence of the phantom. Notice that due to the presence of high load (phantom), resonance frequency of the antennas shifted to the lower frequency.
	
	\begin{figure}
	\centering{\includegraphics[width=98mm]{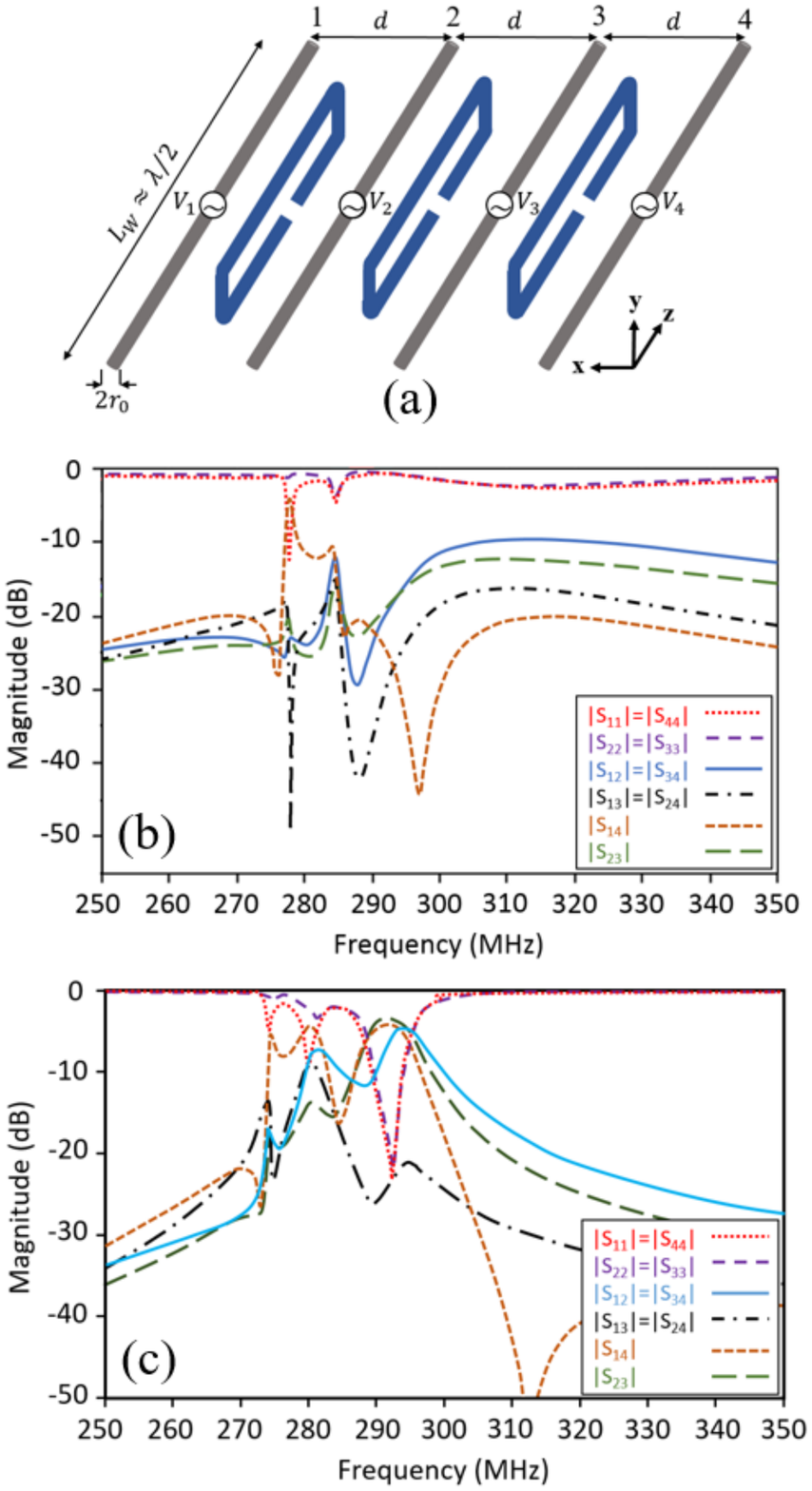}}
	\caption{Array constituted of four active dipoles decoupled by three passive SLRs. (a) Schematic view. (b) Simulation result in mismatched regime. (c) Simulation result in matched regime.}\label{fig5}
\end{figure}

Moreover, in order to emphasize on the possibility of extending this method to arrays with more active dipoles, we simulate the array constituted of four active dipoles decoupled by three SLRs (shown in figure \ref{fig5}(a)). The performance of the structure by the same parameters as before is demonstrated in figures  \ref{fig5}(b) and (c), which shows decoupling between all active dipoles around 290 MHz.
\begin{table}[h!]
	\caption{S-parameter values at 292.3 MHz (dB).}
	\centering
	\begin{tabular}{||c||c|c|c|c||}
		\hline\hline
		Channel & 1 & 2 & 3 & 4 \\ [0.5ex]
		\hline\hline
		1 & -22.1 & -7.1 & -24.8 & -5.3 \\ [0.5ex]
		\hline
		2 & -7.1 & -23.2 & -4.8 & -24.8 \\ [0.5ex]
		\hline
		3 & -24.8 & -4.8  & -23.2 & -7.1 \\ [0.5ex]
		\hline
		4 & -5.3 & -24.8 & -7.1 & -22.1 \\ [0.5ex]
		\hline\hline
	\end{tabular}
\end{table}
In the same manner we can show that the reason of this triple-resonance decoupling band is three different eigenmodes created by three SLRs. For one eigenmode, the induced current over all three are in phase; for second eigenmode, the current over the middle SLR is out of phase and for third eigenmode one of SLRs on the edge is out of phase. We matched this structure at frequency 292.3 MHz. The exact values of S-parameters at this frequency are tabulated in the Table 1. Again, by locating the structure near human body, these values will be enhanced.
These results justifies the possibility of matching for a wider range of frequency and wider decoupled frequency band. This method can be considered as first step for decoupling 32-dipole array used in Ultra-High Field MRI.

	\section{Conclusions}

An approach for creating multi-resonant decoupling band for dense arrays of dipole antennas was discussed. Passive SLRs were located between dipole antennas in order to use their different eigenmodes. The method is based on birefringence of the resonant interaction of SLRs leading to multi-band decoupled performance. Analytical verification of the decoupling was justified by numerical and experimental results. The results prove the double-resonant and triple-resonant of the operation band for array of three and four active dipoles, respectively. For array of three active dipoles, the operation band was enlarged 16 times compared to the previous work. Although the wider operation band was obtained by the price of higher coupling coefficient at the resonant frequency, this wider band makes SLRs an important method for broadening the operation band in dense antenna arrays.
As the main application of the designed structure is 7 T MRI coils, the structure is tuned at 300 MHz. Notice that due to presence of the human body in MRI machine, optimized dimensions of the dipoles and the SLRs as well as the operation frequency may change slightly.

	\section*{Acknowledgment}
	This work was supported by the Russian Science Foundation (Project No. 18-19-00482). Experiments were supported by the European Union's Horizon 2020 research and innovation program under grant agreement No 736937.

\end{document}